\documentstyle[pre,aps,epsf,float,multicol]{revtex}
\begin{document}
\draft
\title{\bf Violation of finite-size scaling for the free energy 
near criticality}
\author{X.S. Chen$^{1,2,3}$ and V. Dohm$^{2}$}
\address{$^1$ Institute of Theoretical Physics, Academia Sinica,
P.O. Box 2735, Beijing 100080, China}
\address{$^2$ Institut f\"{u}r Theoretische Physik, Technische Hochschule
Aachen, D-52056 Aachen, Germany}
\address{$^3$ Institut f\"ur Festk\"orperforschung,
Forschungszentrum J\"ulich, D-52425 J\"ulich, Germany}
\date{\it \today}
\maketitle
\begin{abstract}

The singular part of the finite-size free energy density $f_s$ of the
O$(n)$ symmetric $\varphi^4$ field theory is calculated for confined
geometries of linear size $L$ with periodic boundary conditions in the
large-$n$ limit and with Dirichlet boundary conditions in one-loop
order. We find that both a sharp cutoff and a
subleading long-range interaction cause a non-universal $L$ dependence of
$f_s$ near $T_c$. For film geometry this implies a non-universal critical
Casimir force with an algebraic $L$ dependence that dominates the
exponential finite-size scaling behavior above $T_c$ for both
periodic and Dirichlet boundary conditions.

\end{abstract}
\begin{multicols}{2}

The concept of universal finite-size scaling has played an important role in
the investigation of finite-size effects near critical points over
the last decades \cite{fisher,barber,finite}. Consider the
free-energy density $f (t, L)$ of a finite system at the reduced
temperature $t = (T - T_c) / T_c \geq 0$ and at vanishing external field
in a $d$-dimensional cubic geometry of volume $L^d$ with periodic
boundary conditions (pbc). It is well known that, for small $t$,
the bulk free energy density $f_b \equiv f (t, \infty)$ can be decomposed
as
\begin{equation}
\label{gleichung0} f_b (t) = f_{bs} (t) + f_0 (t)
\end{equation}
where $f_{bs} (t)$ denotes the singular part of $f_b$ and where the
regular part $f_0 (t)$ can be identified unambiguously.
According to Privman and Fisher \cite{privman-fisher,privman} the
singular part of the finite-size free-energy density may be defined by
\begin{equation}
\label{gleichung1} f_s (t, L) = f (t, L) - f_0 (t)
\end{equation}
where $f_0$ is independent of $L$. The finite-size scaling
hypothesis asserts that, below the upper critical dimension $d =
4$ and in the absence of long-range interactions, $f_s (t, L)$
has the structure \cite{privman-fisher,privman,see also}
\begin{equation}
\label{gleichung2} f_s (t, L) = L^{-d} \; {\cal F} (L/\xi)
\end{equation}
where ${\cal F}(x)$ is a universal scaling function and $\xi (t)$ is
the bulk correlation length. Both $\xi$ and $L$ are assumed to be
sufficiently large compared to microscopic lengths (for example, the
lattice spacing $\tilde a$ of lattice models, the inverse cutoff
$\Lambda^{-1}$ of field theories, or the length scale of subleading
long-range interactions). Eq. (\ref{gleichung2}) includes the bulk
limit $f_s (t, \infty) = f_{bs} (t) = Y \xi^{-d}$ with a universal
amplitude $Y$. Eqs. (\ref{gleichung0}) - (\ref{gleichung2})
are expected to remain valid also for non-cubic geometries where the
scaling function ${\cal F} (x)$ depends on the geometry and on the
universality class of the bulk critical point but not on $\tilde a$ or
$\Lambda$ and not on other interaction details 
\cite{privman-fisher,privman,see also,privman1}. In particular subleading 
long-range interactions
(such as van der Waals forces in fluids) that do not affect the universal
bulk critical behavior are assumed to yield only negligible
finite-size corrections to ${\cal F} (x)$.

As a consequence, universal finite-size scaling properties are generally 
believed to hold for observable quantities derived from $f_s(t,L)$, such as 
the critical Casimir force $F_{C}$ in film geometry 
\cite{fisher-gennes,krech,krech1,brankov,krech-dietrich}
\begin{equation}
\label{gleichung31-1} F_{C} = - \partial f^{ex} (t, L)/ \partial L
\end{equation}
where the excess free energy per unit area is given
by
\begin{equation}
\label{gleichung31-2} f^{ex} (t, L) = L f
(t, L) - L f_b (t)\;.
\end{equation}
Near bulk criticality Eqs. (\ref{gleichung0})
- (\ref{gleichung31-2}) yield
\begin{equation}
\label{gleichung21} F_{C} (\xi, L) =  L^{-d} \; X_{C} (L/\xi)
\end{equation}
where
\begin{equation}
\label{gleichung22} X_{C} (x) = (d-1) {\cal F} (x)
- x \; {\cal F}' (x) + Y x^d
\end{equation}
with ${\cal F}' (x) = \partial {\cal F} (x) / \partial x$.
The universal scaling structure of Eqs. (\ref{gleichung21}) and
(\ref{gleichung22}) has been confirmed by renormalization-group (RG) and
model calculations \cite{krech-dietrich,danchev,borjan} and
is predicted to be valid also for Dirichlet boundary conditions (Dbc)
\cite{krech-dietrich}.

In this Letter we show that the scaling predictions of Eqs. (\ref{gleichung2})
and (\ref{gleichung21}) are significantly less generally valid than
anticipated previously. On the basis of exact results for the O$(n)$ symmetric
$\varphi^4$ field theory in the large-$n$ limit and on the basis
of one-loop results we shall identify two sources in the $\varphi^4$
Hamiltonian that cause non-scaling finite-size effects:
(i) a short-range interaction term $\sim {\bf k}^2$ with a {\it sharp}
cutoff $\Lambda$ in ${\bf k}$ space, (ii) an additional subleading
long-range interaction term $\sim b |{\bf k}|^\sigma, 2 < \sigma < 4$.
In the latter case we consider both pbc and Dbc. Specifically, we find
for $2 < d < 4$ that Eqs. (\ref{gleichung2}) and (\ref{gleichung21})
must be complemented as
\begin{equation}
\label{gleichung3} f_s (t, L, \Lambda) \; = \; L^{-2} \; \Lambda^{d-2} \; \Phi
(\xi^{-1} \Lambda^{- 1}) \; + \; L^{-d} \; {\cal F} (L / \xi)
\end{equation}
for the case (i), and
\begin{equation}
\label{gleichung3a} F_{C} (\xi, L, b) \; = \; - b \; L^{- d + 2
- \sigma} \; B (L / \xi)+  L^{-d} \; X_{C} (L / \xi)
\end{equation}
for the case (ii), respectively, where the function
$\Phi$ has a finite critical value $\Phi (0) > 0$ and where the
function $B (L/\xi)$ has a {\it non-exponential}
decay $\sim (L/\xi)^{-2}$ above $T_c$ for both pbc and Dbc. This
implies (i) that the non-scaling $L^{-2}$ term in Eq. (\ref{gleichung3})
exhibits a dominant size dependence compared to the $L^{-d}$ scaling term
and (ii) that the non-universal term proportional to $b$ in Eq.
(\ref{gleichung3a}) implies an algebraic $L$ dependence 
$\sim b \xi^2 L^{-d-\sigma}$ that
dominates the {\it exponential} finite-size scaling term of
$X_{C}$ above $T_c$ for both pbc and Dbc. By contrast, for the
$\varphi^4$ {\it lattice} model with short-range interaction, we find
that Eqs. (\ref{gleichung2}) and (\ref{gleichung21}) are indeed valid 
except that for $L \gg
\xi$ above $T_c$ the exponential scaling arguments of ${\cal F}$
and $X_{C}$ must be formulated in terms of the lattice-dependent 
"exponential" correlation length \cite{chen-dohm-2000,fisher-burford}.

Non-negligible cutoff effects \cite{chen-dohm-1999,chen-dohm-99}
and non-universal finite-size effects due to subleading long-range
interactions \cite{dantchev} were already found previously for the
finite-size susceptibility. These effects, however, were restricted
to the regime $L \gg \xi$ close to the bulk limit above $T_c$.
The new non-scaling finite-size effect (i) exhibited in Eq.
(\ref{gleichung3}) is significantly more general in that it is
pertinent to the entire $\xi^{-1} - L^{-1}$ plane including
the central finite-size regime $\xi \gg L$. In particular,
this effect exists at $T_c$ where $f_s (0, L, \Lambda)$
has a non-universal leading amplitude $L^{-2} \Lambda^{d-2} \; \Phi (0)$
for $d > 2$. This implies the non-universality of the critical Casimir 
force $L^{-2} \Lambda^{d-2} \; \Phi (0)$ in film geometry.

Furthermore, the new non-scaling effect (ii) exhibited in 
Eq. (\ref{gleichung3a}) has relevant physical consequences in systems with
subleading long-range interactions. These
consequences are significantly more important than those
considered previously for the finite-size susceptibility for pbc
\cite{chen-dohm-1999,chen-dohm-99,dantchev}. The latter are of
limited physical relevance since in real systems they are
dominated by surface terms of $O (L^{-1})$. In this Letter
{\it{we predict a leading non-scaling effect on $F_{C}$
not only for model systems with pbc but also for real systems
with Dbc}}.

We start from the standard $\varphi^4$ continuum Hamiltonian
\begin{equation}
\label{gleichung5} H = \int\limits d^d x
\left[\frac{1}{2}\; r_0 \; \varphi^2 \; + \; \frac{1}{2} \; (\nabla
\varphi)^2 + u_0 (\varphi^2)^2 \right]
\end{equation}
with $r_0 = r_{0c} + a_0 t$ for the $n$-component field $\varphi
({\bf x})$ in a partially confined $L^{d'} \times \infty^{d - d'}$
geometry with periodic boundary conditions in $d' < d$ dimensions.
This model requires a specification of the ${\bf x}$ dependence of
$\varphi ({\bf x})$ at short distances. We decompose the vector
$\bf x$ as $({\bf y}, {\bf z})$ where $\bf z$ denotes the
coordinates in the $d'$ confined directions. We consider two
cases (i) and (ii).

{\it Case} (i) : We assume pbc and a sharp cutoff $\Lambda$, i.e., we assume
that the Fourier amplitudes $\hat \varphi_{\bf p,q}$ of $\varphi
({\bf y,z}) = L^{-d'} \sum_{\bf p} \int_{\bf q} \hat \varphi_{\bf p,q}
\;e^{i ({\bf p z} + {\bf q y})}$ are restricted to wave vectors $\bf p$
and $\bf q$ with components $p_j$ and $q_j$ in the range $- \Lambda \leq
p_j < \Lambda$ and $|q_j| \leq \Lambda$. Here $\int_{\bf q}$ stands for
$(2 \pi)^{-d+d'} \int d^{d-d'} q \;$, and $\sum_{\bf p}$ runs over
$p_j = 2 \pi m_j / L$ with $m_j = 0, \pm 1, \pm 2, ...$. The question can
be raised whether there exists a non-negligible cutoff dependence of the
finite-size free energy density per component (divided by $k_B T$),
\begin{equation}
\label{gleichung5a} f_{d,d'} (t, L, \Lambda) = - n^{-1} L^{-d'}
\lim_{\widetilde L \rightarrow \infty} \widetilde L^{-d + d'}
\; \ln Z_{d,d'}
\end{equation}
where
\begin{equation}
\label{gleichung6} Z_{d,d'} (t, L, \widetilde L, \Lambda) = 
\prod_{\bf k,q} \; \int\limits
\frac{d  \;\hat \varphi_{\bf p,q}}{\Lambda^{(d-2)/2}} \; \exp (- H)
\end{equation}
is the dimensionless partition function of a $L^{d'} \times
\widetilde L^{d-d'}$ geometry. For comparison we shall
also consider the free energy density $\hat f (t, L, \tilde a)$ of
the $\varphi^4$ lattice model
\begin{equation}
\label{gleichung7} \hat H = \tilde a^d \left[\sum_i \left(\frac{r_0}{2}
\varphi^2_i + u_0 (\varphi_i^2)^2 \right) \; + \; \sum_{< i j >} \; \frac{J}{2 \tilde a^2} \;
(\varphi_i - \varphi_j)^2 \right]
\end{equation}
with a nearest-neighbor coupling $J$ on a simple-cubic lattice with
a lattice spacing $\tilde a$. The factor $(k_B T)^{-1}$ is absorbed
in $H$ and $\hat H$.

We shall answer this question in the exactly solvable
limit $n \rightarrow \infty$ at fixed $u_0 n$ where the
free energy density is \cite{chen-dohm-529}
\begin{eqnarray}
\label{gleichung15} f_{d,d'} (t, L, \Lambda) &=& - \;
\frac{1}{2} \Lambda^d  \ln \pi \; - \frac{(r_0 - \chi^{-1})^2}
{16 u_0 n} \nonumber\\ 
&+& \frac{1}{2} \; L^{-d'} \sum_{\bf p} \int\limits_q
\ln \left[\Lambda^{-2}(\chi^{-1} + {\bf p}^2 + {\bf q}^2) \right] \; .
\end{eqnarray}
Here $\chi^{-1}$ is determined implicitly by
\begin{equation}
\label{gleichung16} \chi^{-1} = r_0 + 4 u_0 n \; L^{-d'} \sum_{\bf
p} \int\limits_q
\left(\chi^{-1} + {\bf p}^2 + {\bf q}^2 \right)^{-1} \; .
\end{equation}
The bulk free energy $f_b$ and bulk susceptibility $\chi_b$ above
$T_c$ are obtained by the replacement $L^{-d'} \sum_{\bf p} \int_{\bf q}
\rightarrow \int_{\bf k}$, and the critical point is determined by
$r_0 = r_{0c} = - 4 u_0 n \int_{\bf k} {\bf k}^{-2}$ where
$\bf k \equiv ({\bf p}, {\bf q})\;$. The bulk correlation length above
$T_c$ is $\xi = \chi_b^{1/2} = \xi_0 t^{- \nu}$
where $\nu = (d-2)^{-1}$. The regular part of $f_b$ reads $f_0 \; = \; 
\tilde c_1 \; \Lambda^d \;- \; r_0^2 / (16 u_0 n)$ where 
$\tilde c_1$ is a $d$ dependent constant.
The singular part of $f_b$ above $T_c$ is $f_{bs} = Y
\xi^{-d}$ with the universal amplitude 
$Y = (d - 2) A_d / \left[2d (4 - d)\right]$ where
$A_d = 2^{2-d}\pi^{-d/2}(d-2)^{-1}\Gamma (3 - d/2)$.
For the singular part $f_s = f_{d,d'} - f_0$ of the finite-size free
energy above and at $T_c$ we find the form of Eq. (\ref{gleichung3})
with the leading non-scaling part
\begin{eqnarray}
\label{gleichung14} \Phi_{d,d'} (\xi^{-1} \Lambda^{-1}) =
&&\frac{d'}{6 (2 \pi)^{d-2}} \int\limits^\infty_0 d y
\left[\int\limits^1_{-1} d q \; e^{- q^2 y} \right]^{d-1} 
\times \nonumber\\
&\times& \exp
\left[- (1 + \xi^{-2} \Lambda^{-2}) y \right]
\end{eqnarray}
and the subleading universal scaling part
\begin{eqnarray}
\label{gleichung30} {\cal F}_{d,d'} (L/\xi) &=& \;
\frac{A_d}{2 (4-d)} \left[(L/\xi)^{d-2} P^2 - \frac{2}{d} \; P^d
\right] \nonumber\\[12pt] 
&+& \frac{1}{2} \int\limits^\infty_0 \frac{d y}{y} 
\left(\sqrt{\frac{\pi}{y}}\right)^{d-d'} \;
W_{d'}(y) \; e^{- P^2 y / 4 \pi^2}
\end{eqnarray}
where $P (L/\xi)$ is determined implicitly by
\begin{eqnarray}
\label{gleichung31} P^{d-2} = (L/\xi)^{d-2} \; &-& \;
\frac{4-d}{4 \pi^2 A_d} \int\limits^\infty_0 d y \;
\left(\sqrt{\frac{\pi}{y}}\right)^{d-d'} \times \nonumber \\
&\times& W_{d'} (y) e^{-P^2 y / 4 \pi^2} \; ,
\end{eqnarray}
\begin{equation}
\label{gleichung13a} W_d (y) = \left(\sqrt{\frac{\pi}{y} }\right)^{d}
\; - \; \left(\sum^\infty_{m = - \infty} e^{- y m^2}\right)^d \; .
\end{equation}
This result remains valid also for $t < 0$ after replacing the terms
$(L/\xi)^{d-2}$ in Eqs. (\ref{gleichung30}) and (\ref{gleichung31})
by $t (L/\xi_0)^{d-2}$ and after dropping the term $- \xi^{-2}
\Lambda^{-2} y$ in the exponent of Eq. (\ref{gleichung14}).
We have confirmed the structure of Eq. (\ref{gleichung3}) also for
the $\varphi^4$ theory with {\it finite} $n$ within a one-loop
RG calculation at finite $\Lambda$ which yields
the same form of the function $\Phi_{d,d'} (\xi^{-1} \Lambda^{-1})$ as in Eq.
(\ref{gleichung14}). Thus the universal scaling form
(\ref{gleichung2}) is invalid for the $\varphi^4$ field theory with pbc
and with a sharp cutoff, both for $T \geq T_c$ and for $T < T_c$.

These results have a significant consequence for the critical
Casimir effect. Instead of Eq. (\ref{gleichung21}) we obtain from
Eqs. (\ref{gleichung3}) and (\ref{gleichung14}) -
(\ref{gleichung13a}) in film geometry $(d' = 1)$
\begin{equation}
\label{gleichung23a} F_{C} (\xi, L, \Lambda) = L^{-2}
\Lambda^{-2} \; \Phi_{d,1} \; (\xi^{-1} \Lambda^{-1}) 
+ L^{-d} \; X_{C} (L/\xi) \; .
\end{equation}
Thus the critical Casimir force has a {\it leading} non-universal term
$\sim L^{-2}$, in addition to the {\it subleading} universal terms
$\sim L^{-d}$ of previous theories
\cite{krech,krech1,brankov,krech-dietrich,danchev},
both for $T \geq T_c$ and for $T < T_c$.

We have also calculated $f_{d,d'}$ and $F_{C}$ for the
lattice Hamiltonian (\ref{gleichung7}) and for the continuum
Hamiltonian (\ref{gleichung5}) with a {\it smooth} cutoff in the
large-$n$ limit. In both cases the scaling form (\ref{gleichung2})
is found to be valid. For the lattice model, however, the
second-moment bulk correlation length $\xi$ in the argument of
${\cal F}$ must be replaced by the lattice-dependent exponential
correlation length \cite{chen-dohm-2000,fisher-burford}. Specifically
we find, at fixed $t > 0$, the exponential large-$L$ behavior
\begin{equation}
\label{gleichung25} f_s (t, L, \tilde a) - f_{bs} =
- d' (L / 2 \pi \xi_1)^{(d-1)/2} \;L^{-d} \exp (- L / \xi_1)
\end{equation}
where $\xi_1 = (\tilde a / 2) \;  [{\rm arcsinh} (\tilde a / 2 \xi)
]^{-1}$ is the exponential correlation length in the direction of
one of the cubic axes. Note that the {\it non-universal} dependence 
of $\xi_1$ on
$\tilde a$ is non-negligible in the exponent of (\ref{gleichung25})
\cite{chen-dohm-2000}.

The sensitivity of $f_s (t, L, \Lambda)$ and $F_{C}$ with respect
to the cutoff procedure can be explained in terms of a corresponding 
sensitivity of the
{\it bulk} correlation function $G ({\bf x}) = \; < \varphi ({\bf x})
\varphi (0) >$ in the range $|{\bf x}| \gg \xi$ \cite{chen-dohm-2000}.
For example, for the $\varphi^4$ continuum Hamiltonian (\ref{gleichung5}) with
an isotropic sharp cutoff $|{\bf k}| \leq \Lambda$ we find, in the large-$n$
limit, the oscillatory power-law decay above $T_c$
\begin{eqnarray}
\label{gleichung24} G ({\bf x}) = &2 & \Lambda^{d-2} 
(2 \pi x \Lambda)^{- (d+1)/2}
\;  \; \frac{\sin \; [\Lambda x - \pi (d-1) /
4]}{1 + \xi^{-2} \Lambda^{-2}} \nonumber \\
&+& O \left(e^{- x / \xi}\right)
\end{eqnarray}
for large $x = |{\bf x}| \gg \xi$ corresponding to the existence of
long-range spatial correlations which dominate the exponential
scaling dependence $\sim e^{- x / \xi}$.
By contrast, $G ({\bf x})$ has an exponential decay for the lattice model
(\ref{gleichung7}) with purely short-range interaction
\cite{chen-dohm-2000}. An exponential decay of $G ({\bf x})$ is
also valid for the continuum model (\ref{gleichung5}) with a
smooth cutoff \cite{chen-dohm-2000}.

The non-universal cutoff effects on $f_s$, $F_{C}$ and
on $G (\bf x)$ described above are a consequence of the long-range
correlations induced by the sharp-cutoff procedure in the presence
of pbc. We consider these consequences not only as a mathematical
artifact but rather as an important signal for a serious lack of
universality in physical systems. We substantiate this
interpretation by demonstrating that corresponding violations of
finite-size scaling should indeed exist in physical systems with more
realistic interactions and boundary conditions.

{\it Case} (ii) : We assume the existence of a subleading
long-range interaction in the continuum $\varphi^4$ Hamiltonian
$H$ which in the Fourier representation has the form $b |{\bf
k}|^\sigma$ with $2 < \sigma < 4$, in addition to the short-range
term ${\bf k}^2$. It is well known that the subleading interaction
$\sim |{\bf k}|^\sigma$ corresponds to a spatial interaction
potential $V ({\bf x}) \sim |{\bf x}|^{-d-\sigma}$ that
does not change the universal bulk critical behavior \cite{fisher-ma}.
Interactions of this type exist in real fluids.
As pointed out by Dantchev and Rudnick \cite{dantchev},
the presence of this interaction yields leading non-scaling finite-size
effects on the susceptibility $\chi$ for the case of pbc
in the regime $L \gg \xi$ above $T_c$, similar to those found for
a sharp cutoff \cite{chen-dohm-1999,chen-dohm-99}.

In real systems with non-periodic boundary conditions, however, these
non-scaling finite-size effects become only subleading corrections that
are dominated by the surface terms of $\chi$ of $O (L^{-1})$. In the
following we show that the situation is fundamentally different for
$F_{C}$ which, by definition, does not contain contributions of
$O (L^{-1})$ arising from the $O (L^{-1})$ part $\tilde{f}$ of the
free energy density.

We consider film geometry and first assume Dbc in the $z$
direction corresponding to $\varphi ({\bf y}, 0) = \varphi ({\bf
y}, L) = 0$, i.e., we assume that
$\Sigma_p$ in the Fourier representation of $\varphi ({\bf y}, z)
= L^{-1} \sum_p \int_{\bf q} \hat \varphi_{p,{\bf q}} \; e^{i{\bf q
y}} \sin (pz)$ runs over $p = \pi m / L, m = 1,2,...$. Such boundary conditions
are relevant for the superfluid transition of $^4$He \cite{dohm}.
The presence of the subleading interaction $b |{\bf k}|^\sigma$ implies,
for $L \gg \xi$ above $T_c$, a non-universal term $\sim b$ in
\begin{equation}
\label{gleichung26a} f_s (t, L, b) - \tilde{f} \; = \; - b L^{-d
+ 2 - \sigma} \; \Psi (L/\xi) \; + \; L^{-d} \; {\cal G} (L/\xi)
\end{equation}
where ${\cal G} (L/\xi)$ is the known \cite{krech-dietrich} universal 
scaling function
for purely short-range interaction. The scaling function ${\cal G}
(L/\xi)$ has an {\it exponential} large-$L$ behavior
\cite{krech-dietrich}. By contrast we find that $\Psi (L/\xi)$ has
an algebraic $L$ dependence. Performing a one-loop RG calculation we obtain
\begin{equation}
\label{gleichung24a} \Psi (L/\xi) = \frac{1}{2} \; (2 \pi)^{\sigma
- 4} \int\limits^\infty_{(L/\xi)^2} dx \left(1 + x
\frac{\partial}{\partial x} \right) \widetilde \Psi (x) \;,
\end{equation}
\begin{eqnarray}
\label{gleichung24b} \widetilde \Psi (x) =
\int\limits_0^\infty &dy& \; y^{(2-\sigma)/2} \;
e^{-x y / 4 \pi^2} \left(\sqrt{\frac{\pi}{y}} \right)^{d-1}\; 
\widetilde W_{1} (y) \times \nonumber \\
&\times& \gamma^* \left(\frac{2-\sigma}{2}\;, 
- \frac{x y}{4 \pi^2} \right)
\end{eqnarray}
where $\gamma^* (z, x) = x^{- z} \; \int_0^x \; dt e^{-t} \; t^{z-1} 
/\int_0^\infty \; dt e^{- t} \; t^{z-1}$
is the incomplete Gamma function and
\begin{equation}
\widetilde W_1 (y) \; = \sqrt{\frac{\pi}{y}}
\; - \; \frac{1}{2} \sum^\infty_{n = - \infty} \; \exp \left(- \;
\frac{y}{4} \; n^2 \right) .
\end{equation}
We have found that cutoff effects are negligible for the function 
$\Psi (L/\xi)$. At fixed $\xi$, the large-$L$ behavior is 
$\Psi (L/\xi) \sim (L/\xi)^{-2}$.
Eq. (\ref{gleichung26a}) yields the following form
\begin{equation}
\label{gleichung28} B(L/\xi) \; = \; (d - 3 + \sigma) \; \Psi (L/\xi) -
(L/\xi) \; \Psi' (L/\xi)
\end{equation}
for the non-universal contribution to $F_{C}$ in Eq. (\ref{gleichung3a}).
The crucial consequence is that the leading critical temperature dependence
$\sim b \xi^2 L^{-d-\sigma}$ of $F_{C}$ for $L \gtrsim \xi$ above $T_c$
is algebraic and non-universal whereas the critical temperature
dependence of the scaling part $X_{C} (L/\xi)$ derived from ${\cal G}(L/\xi)$ 
is exponential and 
universal \cite{krech-dietrich}. This may have significant consequences for
the interpretation of existing \cite{garcia} and future experimental
data in fluids.

For comparison we finally present our result for $\Psi (L/\xi)$
for film geometry in the presence of pbc. In one-loop order we obtain 
for $\Psi_{pbc}
(L/\xi)$ the same form as given for $\Psi (L/\xi)$ in Eqs.
(\ref{gleichung24a}) and (\ref{gleichung24b}) but with $\widetilde W_1
(y)$ replaced by $W_1 (y)$, Eq. (\ref{gleichung13a}). For the
large-$L$ behavior we find $\Psi_{pbc} (L/\xi) \sim (L/\xi)^{-2}$
which dominates the exponential scaling dependence of
$X_{C}$. Our prediction of a non-exponential non-scaling
effect on $F_{C}$ above $T_c$ for pbc can be tested by Monte Carlo
simulations \cite{krech1}.

The next stage of the theory would entail a quantitative
determination of the non-universal interaction parameter $b$ for a
specific system. This is, of course, beyond the scope of the
present paper.

We thank the referees for stimulating comments.
Support by DLR and by NASA under contract numbers 50WM9911 and
1226553 is acknowledged.

\end{multicols}
\end{document}